\documentclass[aps,prb,
 amsmath,amssymb,
 reprint,%
]{revtex4-2}

\usepackage{graphicx}
\usepackage{dcolumn}
\usepackage{bm}
\usepackage{xcolor}
\usepackage{booktabs}
\graphicspath{{./figures/}}

\usepackage{hyperref}
\hypersetup{
    colorlinks=true,  
    linkcolor=blue,   
    citecolor=blue,  
    urlcolor=cyan     
}
\usepackage[utf8]{inputenc}
\usepackage[T1]{fontenc}
\usepackage{etoolbox}
\usepackage{newtxtext}
\usepackage{newtxmath}
\makeatletter
\def\@email#1#2{%
 \endgroup
 \patchcmd{\titleblock@produce}
  {\frontmatter@RRAPformat}
  {\frontmatter@RRAPformat{\produce@RRAP{*#1\href{mailto:#2}{#2}}}\frontmatter@RRAPformat}
  {}{}
}%
\makeatother
\begin{document}

\preprint{AIP/123-QED}

\title{First-Principles Calculations of the Electron-Hole-Pair Creation Energy in Semiconductors and Dielectrics}
\author{Dallin O. Nielsen}
\email{dallin.nielsen@utdallas.edu}
\affiliation{Department of Materials Science and Engineering, The University of Texas at Dallas, 800 W. Campbell Rd., Richardson,
Texas 75080, USA}%

\author{Massimo V. Fischetti}%
\affiliation{Department of Materials Science and Engineering, The University of Texas at Dallas, 800 W. Campbell Rd., Richardson,
Texas 75080, USA}%

\date{\today}

\begin{abstract}
We present first-principles calculations of the electron-hole-pair creation energy in semiconductors and dielectrics, focusing in particular on AlN, 3C-SiC, 4H-SiC, Si, and SiO$_2$. We have employed density functional theory (DFT) to calculate the band structure, density-functional perturbation theory to obtain the phonon dispersion, and time-dependent DFT (TDDFT) to calculate the electron energy loss function (via the dielectric function). We performed full-band Monte Carlo simulations of electron thermalization for energies up to $\sim100$~eV and calculated the average energy required to excite an electron-hole pair. We calculated pair-creation energies of approximately 12.2, 6.5, 7.3, 3.10, and 16.7~eV/pair for AlN, 3C-SiC, 4H-SiC, Si, and SiO$_2$, respectively. These values are in good agreement with measured data, with the largest discrepancy being that of Si (about 14.9\% lower than the accepted experimental value of $\sim$~3.6 eV/pair). To improve the agreement for Si, we used the GW approach and the Bethe-Salpeter equation (BSE) to calculate the quasiparticle shifts of the electronic band structure and the dielectric function, respectively. Finally, using the GW/BSE data, we calculated a creation energy of $\sim3.50$~eV/pair. We find that the marked improvement in the accuracy of the calculated creation energy, when using the GW/BSE approach, is primarily attributable to the slight decrease in the magnitude of the electron energy-loss function when using the GW/BSE approach for energies below $\sim$~10-15~eV. Overall, it is clear that the competition between phonon emission and ionization near the intersection of the phonon scattering and ionization rates is largely responsible for determining the value of the creation energy, with the other major factor being the location and shape of the plasmon peak in the energy-loss function.
\end{abstract}

\maketitle

\section{\label{sec:Introduction}Introduction}

When ionizing radiation interacts with a semiconductor or dielectric, it deposits energy mainly by generating electron-hole pairs (EHPs) and, in the case of ions, interacting with the material nuclei. Knowledge of the average energy required to excite or generate an EHP is important for the study of the stopping power and understanding the radiation resistance of a given material and/or device, as well as accurately simulating radiation effects. It is also important in the study of radiation detection~\cite{shrestha2017,liu2020,fano1947,david2016}, photovoltaics~\cite{polman2016}, radioisotope microbatteries~\cite{dixon2016,olsen2012}, and optoelectronics~\cite{arp2019}. This quantity is commonly referred to as the EHP creation energy or radiation-ionization energy~\cite{fang19}, typically denoted in the literature by the symbols $\epsilon$ or $W$. In this paper, we will use the symbol $\epsilon$ to represent the creation energy, following the notation used by McKay~\cite{mckay1951} and Alig,~{\it et al.}~\cite{alig75}. 

In 2023, we published two papers~\cite{nielsen23a,nielsen23b}, in which we studied, using density functional theory (DFT), the physics of electronic transport in the energy regime of 10-100 eV, an area of study that was previously poorly understood. Following the work of Pines~\cite{pines56} and others~\cite{rothwarf1973,ausman1975,kingsley1970}, we incorporated plasmon emission, impact ionization, and phonon scattering into a full-band Monte Carlo (FBMC) code to simulate carrier thermalization from $\sim100$~eV. Though these papers were primarily focused on developing the physics and understanding the thermalization process, a key result was the calculation of $\epsilon$. For GaN, diamond, and $\beta$-Ga$_2$O$_3$, we calculated values of 8.9, 12.87, and 11.24~eV~\cite{nielsen23a,nielsen23b}, respectively. In the present work, we have extended the techniques used in our past works to AlN, 3C-SiC, 4H-SiC, Si and SiO$_2$, using not only DFT but also the GW approach and the Bethe-Salpeter equation (BSE), focusing now on the pair creation energy rather than on the dynamics of carrier thermalization. The listed materials have band gaps of 6.2, 2.36, 3.2, 1.12, and 9.0~eV, respectively. This relatively wide range of band gap widths allows for a comprehensive assessment of the predictive power of the first-principles approach in calculating $\epsilon$ and its dependence on $E_g$ -- a relationship that has been studied since the 1940s, due to its importance in radiation detection and radiation damage. 
\begin{figure}
\includegraphics[height=3.25in, width=3.37in]{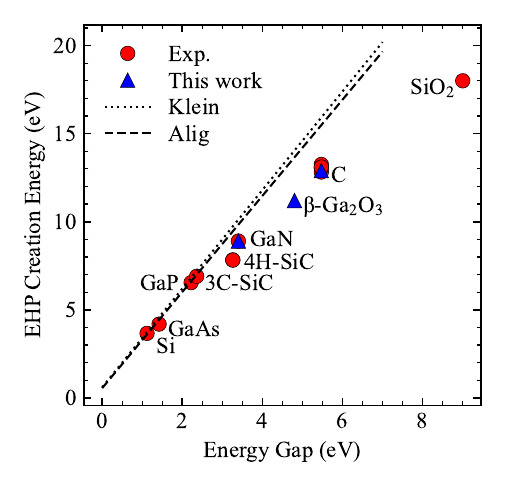}
\caption{\label{fig:ehp_creationen_exp} Measured EHP creation energies (red circles) for several materials, including Si~\cite{scholze2000,lowe2007}, GaAs~\cite{bertuccio2002}, GaP~\cite{kobayashi1972}, 3C-SiC~\cite{makarov1975a,makarov1975b}, 4H-SiC~\cite{garcia2013,gsponer2024}, GaN~\cite{sellin06}, and diamond (C)~\cite{kozlov1975,canali1979,keister2009,morse2007,keister2018}, compared to the analytic expressions of Klein~\cite{klein68} and Alig~{\it et al.}~\cite{alig75}.}
\end{figure}


Bethe developed early analytic approximations of the stopping power of ions in matter that included $\epsilon$ as a parameter~\cite{bethe30}, leading to an increased demand for accurate creation energy data to improve radiation damage predictions. Measurements of $\epsilon$ began in the 1940s and 1950s with the works of Ahearn~\cite{ahearn48}, McKay~\cite{mckay1951}, and McKay and McAfee~\cite{mckay1953}. Lappe was the first to note that the reported values of $\epsilon$ were approximately equal to three times the bandgap energy\cite{lappe61}, $E_{\rm g}$, a simplification commonly still assumed today. Baldinger and Czaja suggested that the relationship between $\epsilon$ and $E_{\rm g}$ is of the form $\epsilon=k_1E_{\rm g}+k_2$, a correlation similar to that for gases~\cite{baldinger1961}, leading to many attempts to develop such a linear relationship. Some notable examples include those of Shockley~\cite{shockley61}, Klein~\cite{klein68}, and Alig and Bloom~\cite{alig75}.  

In Fig.~\ref{fig:ehp_creationen_exp}, the analytic expressions of Klein~\cite{klein68} and Alig~{\it et al.}~\cite{alig75} are shown along with the measured values of $\epsilon$ (red circles) for various materials and values that we calculated in past publications~\cite{nielsen23a,nielsen23b} (blue triangles), which we mentioned above. It is clear from the figure that both analytic expressions agree well with the measured data for materials with E$_g\lesssim3$~eV. For larger bandgaps, the measured creation energies diverge from these analytical predictions as the measured values consistently fall below the two lines. This deviation for wide bandgap materials has led to more recent developments of a new analytic expression~\cite{whitaker2020,lioliou2022} with a smaller slope, capturing the data for larger $E_{\rm g}$ better at the expense of the fit for small values of $E_{\rm g}$. However, it is clear that the slope of the measured data is not consistent over the entire domain of bandgap energies shown in Fig.~\ref{fig:ehp_creationen_exp}. Indeed, the divergence seems to increase with increasing bandgap, indicating that, while the qualitative trend is clear, the quantitative relationship between $\epsilon$ and $E_{\rm g}$ is not exactly linear. The agreement between our calculated results and the measured data for GaN and diamond, shown in Fig.~\ref{fig:ehp_creationen_exp}, suggests that by employing a first-principles approach with a FBMC simulation, one may accurately predict this deviation from the linear behavior. 

The present paper is outlined as follows: In Sec.~\ref{sec:past_approaches}, we will discuss in more detail the methods used in past works, especially those of Shockley~\cite{shockley61} and Alig, Bloom, and Struck\cite{alig75,alig1980}, along with some lingering questions from their works, which we seek to answer here. We will also discuss why the common linear approximations fail and why a physically more accurate calculation is necessary. In Sec.~\ref{sec:pbe_approach}, we will show the calculations/simulations for the above-mentioned materials, employing a standard DFT approach, utilizing  PBE~\cite{perdew96} exchange-correlation functionals. In Sec.~\ref{sec:gwbse_approach}, to improve the predicted $\epsilon$ of Si, we will show an alternative calculation of the band structure and the dielectric function, employing the GW approach to calculate the quasi-particle energy shifts in the band structure and the Bethe-Salpeter equation (BSE), which captures the effect of electron-hole interactions, to calculate the dielectric function (the ``GW/BSE approach''). Lastly, in Sec.~\ref{sec:discussion}, we will discuss the results and answer a few questions posed by Alig~{\it ert al.}.



\section{\label{sec:past_approaches}An Analysis of Past Theoretical Approaches}

As mentioned above, a common and longstanding approach to predicting the creation energy has been to develop a linear relationship between $\epsilon$ and the bandgap energy, $E_{\rm g}$. Early on, in the 1950s, this was done simply by fitting a line to experimentally measured data points without a physical explanation for the fit. Below, we will first discuss in some detail the work done by Shockley in 1961~\cite{shockley61} to attempt to provide such a physical explanation. Afterward, we will discuss the work of Alig, Bloom, and Struck in 1980~\cite{alig1980}, who finally moved beyond the assumption of a simple linear relationship. Lastly, we will touch on the more modern MC simulations and why the present work is necessary in the advancement of the state of the art. We will present some questions that remain from the past, which we will attempt to answer in the remaining sections of this paper.

In Shockley's work, he emphasized the importance of the ``competition'' between impact ionization and phonon emission during the energy-loss process in determining the value of $\epsilon$. When a particle or electromagnetic wave excites an EHP, the excited carriers, if given sufficient energy, will dissipate their energy by causing additional ionization and/or by emitting phonons. After dropping below a certain threshold energy (i.e., below about $ E _ {\rm g} $), ionization is no longer possible, and phonon emission becomes the dominant energy-loss process. Shockley, therefore, described $\epsilon$ in terms of the threshold energy, $E_ i$, the phonon energy, $E_R$, the ratio, $r$, of the phonon scattering rate to the ionization rate for carriers with $E>E_i$, and the average carrier energy, $E_f$, after falling below $E_i$. Following this approach, the general equation for $\epsilon$ that he obtained is~\cite{shockley61}
\begin{equation}
\label{eq:shockley_eps}
\epsilon=2E_f + E_i + rE_R.
\end{equation}

If one employs free-electron bands and assumes that the states below $E_i$ are ultimately uniformly populated (the ``uniform population'' assumption~\cite{alig1980}), one obtains a value of $E_f=0.6E_i$. The equation for the creation energy then becomes $\epsilon=2.2E_i+rE_R$. The common approach is to substitute the longitudinal optical (LO) phonon energy for $E_R$, assuming that scattering with other vibrational modes is negligible at energies above $E_i$~\cite{fischetti2019}. The parameter $r$ then becomes a fitting parameter, assumed to be constant in energy. Using similar approaches, Klein~\cite{klein68} and Alig and Bloom~\cite{alig75} developed the following equations, $\epsilon=(14/5)E_{\rm g}+0.5$ and $\epsilon=2.73E_{\rm g}+0.55$, respectively (shown in Fig.~\ref{fig:ehp_creationen_exp}). As measurements of $\epsilon$ have improved over the years, especially for wide bandgap materials, more recent work~\cite{whitaker2020,lioliou2022} has yielded the following equation $\epsilon=1.54E_{\rm g}+1.89$. In this newer expression, we see a significant reduction in the slope of the line, due mainly to the divergence shown above in Fig.~\ref{fig:ehp_creationen_exp} of $\epsilon$ from published analytic expressions, as $E_{\rm g}$ increases. The issue, as described in the Introduction, is that the relationship between $\epsilon$ and $E_g$ is not linear, and attempting to fit a linear curve to the data on one end of the domain of bandgap widths leads to a poor fit on the other end. 

While Shockley's approach cannot account for the measured nonlinear behaviour of $\epsilon$ as a function of $E_{\rm g}$, it did provide valuable insights. Firstly, it emphasized the importance of the ratio of the phonon emission rate to the ionization rate, which directly led to the work of Alig, Bloom, and Struck (discussed below), whose approach centered on calculating $\epsilon$ by determining the probability of generating an EHP via the rate ratio. Secondly, it introduced the parameter $ E_f$ to the discussion and raised the question of how the states below $E_i$ are ultimately populated.

Alig~{\it et al.} sought to shed some light on the subject by utilizing a new approach which they called the ``scattering-rate'' assumption~\cite{roosbroeck1965,antoncik1968,drummond1971,alig1980} with what they called the ``probability method''. Instead of assuming that ultimately the states below $E_i$ are uniformly populated, the scattering rate assumption states that in each scattering event all possible sets of the product particles are created with equal probability. In other words, all nonzero matrix elements for each scattering mechanism (i.e., phonon scattering and impact ionization) are identical and energy-independent. In practice, this means that, when using Fermi's golden rule to calculate the scattering rates, one may move the matrix elements outside the summation over final states. As an example, Alig approximated the equation for the electron-phonon scattering rates, $r^{({\rm ph})}(E)$ as follows:
\begin{align}
\label{eq:eph_fermi}
r^{({\rm ph})}(E) &= \frac{2\pi}{\hbar}\sum_j \lvert M_j^{({\rm ph})} \rvert^{2} \delta (E-E_j\pm\hbar\omega) \nonumber \\ &\approx \frac{2\pi}{\hbar}\lvert M^{({\rm ph})} \rvert^{2}\sum_j \delta (E-E_j\pm\hbar\omega).
\end{align}
Here, $E$ is the energy of the carrier, $M_j^{({\rm ph})}$ is the matrix element for scattering to a state $j$ of energy $E_j$, and $\hbar\omega$ is the phonon energy absorbed ($+$) or emitted ($-$) in the process. The summation over the delta function simply becomes the total density of states. Alig~{\it et al.} assumed parabolic, isotropic bands within the free electron model.

The ``probability method'' involves calculating the probabilities, $p_n(E)$, of a primary particle of energy $E$ generating exactly $n$ EHPs during the thermalization process. Having calculated $p_n(E)$, one may then calculate the average number of generated pairs using the following equation,
\begin{equation}
\label{eq:avg_n}
\langle n \rangle=\sum_0^\infty np_n(E),
\end{equation}
and, finally, the creation energy can be calculated as
\begin{equation}
\label{eq:creation_en}
\epsilon=E/\langle n \rangle.
\end{equation}
We will not include all of the details of the calculation of $p_n(E)$ here, but we direct the interested reader to the paper by Alig~{\it et al.}~\cite{alig1980}.

 A key piece of information needed in the calculation of $p_n(E)$ is the ratio of the phonon emission rate to the ionization rate (Shockley's ``r''). By dividing the equation for the phonon scattering rates (Eq.~\ref{eq:eph_fermi}) by that for impact ionization via the random-k approximation, Alig~{\it et al.} obtained the following:
\begin{equation}
\label{eq:rate_ratio}
\frac{r^{({\rm ph})}(E)}{r^{({\rm ii})}(E)}=A\frac{105}{2\pi}\frac{(E-\hbar\omega_0)^{1/2}}{(E-E_{\rm g})^{7/2}},
\end{equation}
where the parameter A is defined as
\begin{equation}
\label{eq:a_parameter}
A=\frac{\left| M^{(ph)} \right|^2}{\left| M^{(ii)} \right|^2}\frac{32\pi^4}{\Omega_c^3}\left( \frac{\hbar^2}{2m} \right)^3,
\end{equation}
and $\Omega_c$ is the unit cell volume and $\hbar\omega_0$ is the longitudinal optical phonon energy. In practice, $A$ is treated as the single fitting parameter employed in the calculation. Again, we emphasize that according to the ``scattering-rate'' assumption, the matrix elements are constant, and so, $A$ is constant.

Alig~{\it et al.} fit $A$ to a value of 5.2 so that the method reproduced the experimentally measured value of $\epsilon$ for Si. Astonishingly, using the same value of $A$, they also accurately calculated $ \epsilon$ for several other materials such as Ge, AgBr, and 3C-SiC. For others such as SiO$_2$, C, PbO, and CdS, the method was less accurate. The reasons why Alig's approach was successful (or not) will be made evident in later sections of this paper. In analyzing their results, Alig~{\it et al.} raised a couple of important questions that we will attempt to answer in this work: Is the parameter $A$ constant, or nearly so, for most if not all semiconductors and dielectrics? Is it important to account for plasmon emission, or can it be ignored? 



During the 1980s and 1990s, significant advancements were made in simulating electronic transport in semiconductors and solving the semiclassical Boltzmann transport equation (BTE)~\cite{jacoboni83,shichijo1981,tang1983,fischetti88,fischetti95}. These approaches allow one to move beyond the scattering-rate assumption and account for the unequal probability of final-state selection in the scattering process (i.e., unequal and energy-dependent matrix elements). In the following section, we will show the calculations we performed of $\epsilon$, using a first-principles approach with MC techniques to solve the BTE.


\section{\label{sec:pbe_approach} Calculating the EHP Creation Energy}
\subsection{\label{sec:abinitio}DFT and Scattering Rate Calculations}

We utilized the DFT package Quantum ESPRESSO~\cite{giannozzi09} (QE) for all {\it ab initio} calculations shown in this section, employing norm-conserving pseudopotentials~\cite{schlipf15} and PBE~\cite{perdew96} exchange-correlation functionals. We began by performing a relaxation of the unit cells of each material, using the {\it vc-relax} function in QE. The unit cells of AlN and 4H-SiC are hexagonal, belonging to the P6$_3$mc space group, while that of 3C-SiC is zinc-blende with a space group of F-43m, and that of Si is diamond-structured with a space group of  Fd$\bar{3}$m. Finally, for SiO$_2$, we used the $\alpha$-quartz structure with space group P3$_2$21. The resulting lattice constants, following relaxation, are presented in Table~\ref{tab:lattice_constants}. 
\begin{table}
\caption{Lattice constants obtained from relaxations of the crystal structures of AlN, 3C-SiC, 4H-SiC, and Si, using QE.}
\label{tab:lattice_constants}
\setlength{\tabcolsep}{10pt} 
\renewcommand{\arraystretch}{1.2}
\begin{tabular}{c c c}
\toprule
Material&
a$_0$ (\AA) &
c$_0$ (\AA) \\
\midrule
AlN & 3.11 & 4.98\\
3C-SiC & 4.37 & -\\
4H-SiC & 3.09 & 10.13\\
Si & 5.48 & -\\
SiO$_2$ & 5.03 & 5.52\\
\bottomrule
\end{tabular}
\end{table}

Using the relaxed structures, we performed self-consistent field calculations to converge the Kohn-Sham potential, using a uniform 8$\times$8$\times$8 Monkhorst-Pack grid. We then performed non-self-consistent field calculations of the electronic band structures for uniform 40$\times$40$\times$40 meshes of {\bf k} points, spanning the first Brillouin Zone (BZ). Using the symmetry operations of each material, these meshes were reduced to irreducible sets of {\bf k} points spanning the irreducible wedge of each BZ (3234 {\bf k} points for AlN and 4H-SiC, 1661 for 3C-SiC and Si, and 16422 for SiO$_2$). We calculated and stored the band structure data at these irreducible points for all of the valence bands (VBs) and a sufficient number of conduction bands (CBs) to reach kinetic energies exceeding $\sim100$~eV. The number of necessary CBs varied depending on the material. For Si, we used 136 CBs, while for AlN, 3C-SiC, 4H-SiC, and SiO$_2$, we used 192, 68, 284, 326, respectively. These data were then used for interpolation of the band structure at any arbitrary {\bf k} and band, $n$, in the full-band MC simulations described below.

To incorporate carrier-phonon interactions in our simulations, we employed density functional perturbation theory (DFPT) to calculate the phonon dispersion for the same set of irreducible ({\bf q}) points as above and all vibrational modes ($\eta$) (12 modes for AlN, 6 for 3C-SiC, 24 for 4H-SiC, 6 for Si, and 27 for SiO$_2$). As was done for the band structure, these data were stored in tables for interpolation in the MC simulations. 

The carrier-phonon scattering rate between a carrier of wave vector {\bf k} in band $n$ and a phonon of wave vector {\bf q} and branch $\eta$ was calculated using first-order time-dependent perturbation theory (Fermi's golden rule)~\cite{fang19},
\begin{multline}
\label{eq:eph_rate}
\frac{1}{\tau_n({\bf k})} \approx \frac{2\pi}{\hbar}\sum_{n'{\bf q}}|g_{nn'}^\eta({\bf k},{\bf k'})|^2\left(N_{\bf q}+1/2\mp1/2\right) \\
\times\delta\left[E_n({\bf k})-E_{n'}({\bf k'})\pm\hbar\omega_{\bf q}\right],
\end{multline}
with
\begin{equation}
\label{eq:bose-einst}
N_{\bf q}=\frac{1}{e^{(\hbar\omega_{\bf q}/k_BT)}-1}.
\end{equation}
Here, $g_{nn'}^\eta({\bf k},{\bf k'})$ represents the electron-phonon matrix element, ${\bf k'}={\bf k}\pm{\bf q}$ is the ``final'' wave vector of the scattered carrier, and $n'$ is the final band. The upper sign of the plus-minus signs is chosen for absorption, while the lower is chosen for emission. We used Bl{\"o}chl's tetrahedron method~\cite{blochl94} to evaluate the BZ integration of the delta function. The matrix elements, $g_{nn'}^\eta({\bf k},{\bf k'})$, were calculated using the EPW (Electron-Phonon coupling using Wannier functions) code, which is included in the QE package~\cite{giustino07,ponce16}. The matrix elements were calculated on uniform {\bf k} and {\bf q} point grids for all VBs of each material and the first 10, 20, 12, 32, and 36 CBs for Si, AlN, 3C-SiC, 4H-SiC, and SiO$_2$, respectively. As discussed in Ref.~\cite{nielsen23a}, the number of necessary CBs is determined by comparing the total phonon scattering rates with the so-called energy-loss rates (which will be discussed below) to find an energy above which phonon scattering becomes negligible.  

For electrons and holes possessing kinetic energies greater than the band gap, impact ionization must be considered along with phonon scattering. Additionally, for kinetic energies near and exceeding the bulk plasmon energy, valence-electron plasmon emission must also be taken into account. The plasmon energy of a material can be approximated using the free-electron gas model, 
\begin{equation}
\label{eq:plasmon_freq}
\omega_p^2=\frac{e^2n}{\epsilon_0m}.
\end{equation}
Equation~(\ref{eq:plasmon_freq}) yields results of 22.99, 22.99, 22.93, 16.39, and 23.40~eV for AlN, 3C-SiC, 4H-SiC, Si, and SiO$_2$, respectively. These calculations were carried out using the lattice constants obtained from DFT (see Table~\ref{tab:lattice_constants}). As mentioned in the introduction, we calculated the creation energy by simulating the thermalization of electrons with energies up to $\sim100$~eV, and as each of these plasmon energies is well below 100~eV, it was necessary to incorporate plasmon emission as an energy-loss mechanism in the MC simulation. In fact, Pines showed that losses to plasmons are the dominant process in the 10-100~eV energy range~\cite{pines56}.

To calculate the total rate of impact ionization and plasmon emission, we utilized the energy-loss function, or the imaginary part of the inverse dielectric function, Im$\left[ -1/\epsilon({\bf q},\omega) \right]$, as in Ref.~\cite{nielsen23a}. The loss function is essentially a spectrum that indicates which energies are most likely to be lost when a charged particle interacts with the valence electrons in a material, either via single-particle excitations (impact ionization) or collective excitations (plasmon emission). Using Fermi’s golden rule and the dissipation-fluctuation theorem, we obtained an equation to calculate the total rate of both impact ionization and plasmon emission, which we call the energy-loss rate (ELR),
\begin{multline}
\frac{1}{\tau_{n}^{\rm (ELR)}({\bf k})} = 
   \frac{2\pi}{\hbar} \sum_{n'} \int\frac{{\rm d}{\bf q}}{(2\pi)^3} \frac{e^{2}\hbar}{q^2} 
       \int\frac{{\rm d}\omega}{2\pi} \ {\rm Im} \left[\frac{-1}{\varepsilon({\bf q},\omega)}\right] \\
           \times \delta \left [ E_n({\bf k})-E_{n'}({\bf k}+{\bf q}) \pm \hbar\omega \right]. 
\label{eq:ELR}
\end{multline}

To calculate the loss function, we utilized the time-dependent DFT (TDDFT) code {\it turboEELS}~\cite{timrov15}, which is included in the QE package. To begin, we performed a standard self-consistent DFT calculation utilizing a $12\times12\times12$ Monkhorst-Pack grid. The output files of this calculation were read by {\it turboEELS}, which then calculated the dielectric function for a given wavevector, {\bf q}. Additionally, we used 1000 Lanczos iterations with a bi-constant extrapolation of the Lanczos coefficients up to 20000 and a Lorentzian broadening of 0.02 eV~\cite{nielsen23a,nielsen23b,nielsen26a}. As in Ref.~\cite{nielsen23a}, we assumed a spherically isotropic loss function in reciprocal space, allowing us to calculate the dielectric function only along a single reciprocal lattice vector for several {\bf q} points. This approach significantly reduces computational cost, in comparison to a calculation for many {\bf q} in the irreducible wedge, without appreciably affecting the accuracy of the result~\cite{nielsen23a}.

\begin{figure}
\includegraphics[height=3.25in, width=3.37in]{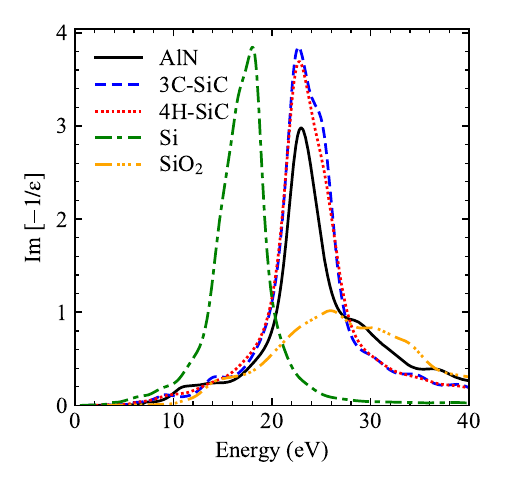}
\caption{\label{fig:loss_functions} The loss functions of AlN, 3C-SiC, 4H-SiC, and Si calculated using the {\it tuboEELS} code, a time-dependent DFT code included in the QE package.}
\end{figure}

The loss functions of each material are shown in Fig.~\ref{fig:loss_functions}. The primary peaks, which represent the primary bulk plasmons of each material, occur at energies of $\sim22.95$, 22.7, 22.7, 17.9, and 25.9~eV for AlN, 3C-SiC, 4H-SiC, Si, and SiO$_2$, respectively. As the calculations of the plasmon energy using the free-electron gas model showed, the peaks of AlN, 3C-SiC, and 4H-SiC are at nearly equivalent locations. Additionally, the shape and magnitude of the curves for 3C-SiC and 4H-SiC are strikingly similar. One may also observe that the plasmon peak of SiO$_2$ is significantly broader and smaller in magnitude than those of the other materials shown in Fig.~\ref{fig:loss_functions}. This indicates that the uncertainty in the energy of the valence-electron plasmons of SiO$_2$ is higher, and, by Heisenberg's uncertainty principle, the lifetime (or coherence time) is shorter. The lifetime is controlled by Landau damping, by which a plasmon decays into an electron-hole pair. The relatively flat and broad nature of the SiO$_2$ peak indicates that the plasmons decay more quickly, relative to those in the other materials. Additionally, the broadness of the peak affects the calculated creation energy. While for a narrow peak, electrons with energies above the peak energy will tend to generate pairs with relatively discrete energies, corresponding to the location of the peak, for broad peaks, the pair energy distribution is more continuous. This makes the energy cascade less predictable, requiring more simulations/samples for better statistics. 

Looking at Eq.~\ref{eq:ELR}, one may also observe that the magnitude of and the area under the loss function curve give a good indication of the trends one may expect to see in the plots of the ELRs. In Fig.~\ref{fig:loss_functions}, up to approximately 28~eV, the magnitude of the loss functions of AlN, 3C-SiC, and 4H-SiC exceed that of SiO$_2$, and it is clear that the areas under the curves, for these three materials, are also larger. One may therefore expect the ELRs of AlN, 3C-SiC, and 4H-SiC to also exceed those of SiO$_2$ over this energy range. The similarities, discussed above, between the loss functions of 3C-SiC and 4H-SiC indicate that the ELRs will likewise be similar. 

\begin{figure}
\includegraphics[height=3.25in, width=3.37in]{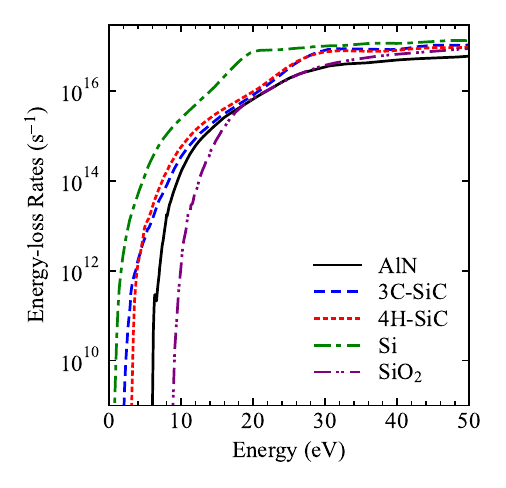}
\caption{\label{fig:elr} The electron ELRs for AlN, 3C-SiC, 4H-SiC, Si, and SiO$_2$.}
\end{figure}

\begin{figure*}
\includegraphics[height=6.0in, width=7.0in]{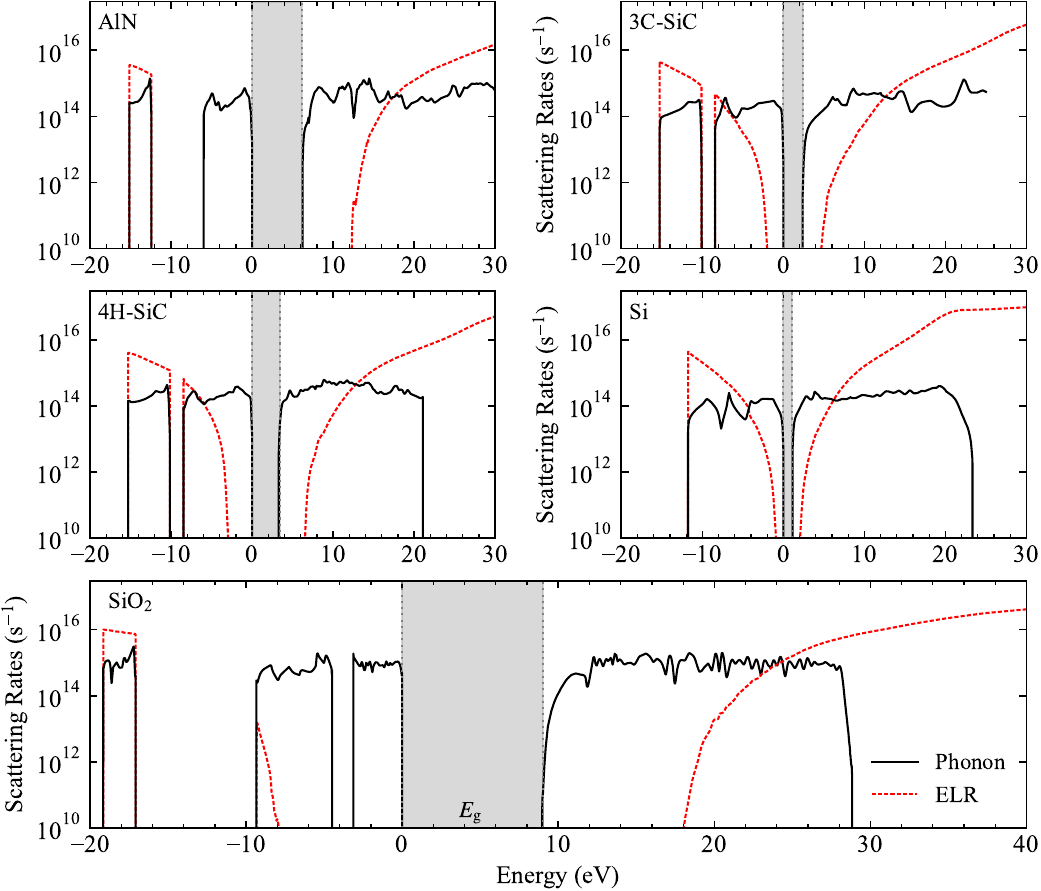}
\caption{\label{fig:scatt_rates} The total phonon scattering rates (black, solid lines) and the ELRs (red dashed lines) for each material (indicated by the label in the upper left corner of each frame). The gray shaded regions represent the direct energy gaps. This is included to help the reader distinguish between hole (left of shaded region) and electron (right of shaded region) scattering rates.}
\end{figure*}

With the loss functions calculated, we evaluated Eq.~\ref{eq:ELR} for each material. The resulting electron ELRs are shown in Fig.~\ref{fig:elr}.
For each material, the ELRs increase rapidly above the bandgap energy and eventually level off near $\sim10^{17}$~s$^{-1}$. As one might expect, based on the analysis of Fig.~\ref{fig:loss_functions}, the ELRs of silicon reach this point of leveling off near about 20~eV, just above the plasmon peak energy. These rates level off rather sharply, due to the narrow nature of the loss function peak of Si. For SiO$_2$, the ELRs climb more gradually, which is caused by the much flatter loss function. As is the case for the loss functions, the ELRs of the two polymorphs of SiC are very similar, nearly lying on top of one another over the entire energy range. We also show, in Fig.~\ref{fig:scatt_rates}, the total phonon scattering rates (black solid line) with the ELRs for each material (red dashed line). This figure includes the rates for both holes and electrons (separated by the gray shaded region, which indicates the energy gap). As mentioned above, the phonon scattering rates are calculated up to kinetic energies above which the ELRs dominate, rendering phonon scattering negligible. 

As discussed in Refs.~\cite{nielsen23a,nielsen23b}, we note that the relatively large ELRs shown in Fig.~\ref{fig:scatt_rates} may raise questions about the validity of first-order time-dependent perturbation theory (Fermi's golden rule). This concern has been addressed by Quinn and Ferrell~\cite{quinn58,quinn62}, Penn~\cite{penn87}, and Pines~\cite{pines56}, who found that the electron kinetic energies at which the ELRs are high are large enough to render the broadening of the states acceptably small, justifying the use of first-order perturbation theory.

\subsection{\label{sec:mc_sim}Monte Carlo Simulations}

With the material properties and scattering rates calculated, we were ready to simulate the thermalization of hot electrons by solving the semi-classical Boltzmann transport equation via full-band MC techniques. Here, we will give only a short overview of the MC simulation, and we direct the interested reader to our previous work~\cite{nielsen23a}, which contains the numerical details.
\begin{figure*}
\includegraphics[height=3.5in, width=6.0in]{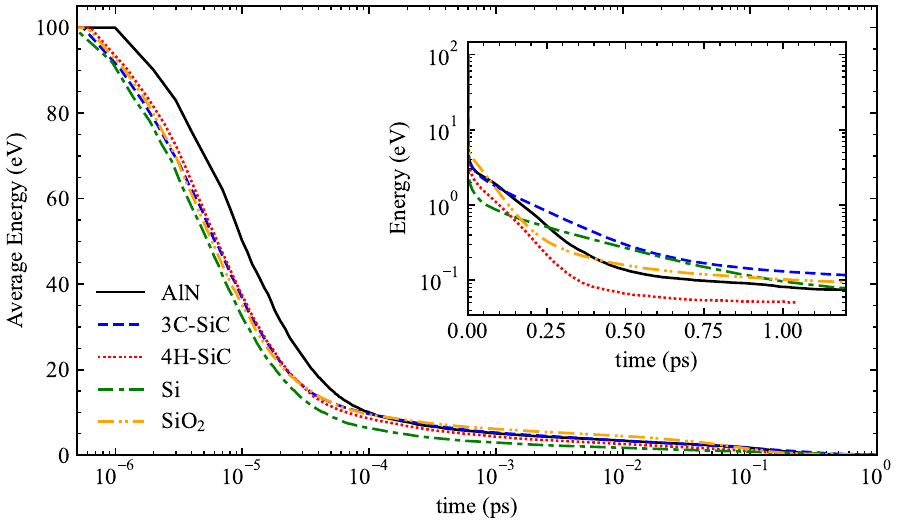}
\caption{\label{fig:therm_rates}The average electron kinetic energy as a function of time for a simulation of the relaxation of 100-eV electrons. The larger figure shows the entire simulation with the time axis on a log scale. The inset shows the same curves but with the energy axis on a log scale so that the reader may better see the later stages of the thermalization process.  }
\end{figure*}

In each material, we initialized a set of electrons (between 1000 and 10000 particles) with kinetic energies chosen under a Gaussian distribution centered at 100~eV (above the CB minimum). Their initial ${\bf k}$ points and bands ($n$) were chosen by discretizing the BZ and randomly selecting a cell on the BZ mesh, for a given band, that contained the kinetic energy of the electron. A ${\bf k}$-point inside the chosen cell was selected from a set of ${\bf k}$-points on a submesh within the cell, such that the difference between the electron energy and the energy at the ${\bf k}$-point was minimized. 

With no applied field, we then simply allowed the electrons to diffuse and scatter via phonon scattering, impact ionization, and plasmon emission. We note that, as in Refs.~\cite{nielsen23a,nielsen23b}, we assumed immediate decay of emitted plasmons via Landau damping, leading to the creation of an EHP. We justified this assumption in Ref.~\cite{nielsen23a} by showing that the plasmon lifetimes are extremely short, being on the order of $\sim10^{-16}$~s. This reduces the complexity of the MC simulation, allowing one to treat impact ionization and plasmon emission as effectively identical processes. 

As described in Refs.~\cite{nielsen23a,nielsen23b}, while the electrons possess kinetic energies above $\sim10$~eV, the dominant energy-loss mechanisms are impact ionization and plasmon emission. As shown in Fig.~\ref{fig:scatt_rates}, the ELRs are significantly higher than the phonon scattering rates in the energy range of approximately 10-100~eV, about two orders of magnitude higher across much of the range. These relatively high ELRs lead to rapid thermalization, and the electrons lose $\sim90\%$ of their kinetic energy within tenths of a femtosecond. For a visualization of this process, see Fig.~\ref{fig:therm_rates}. After this rapid loss of energy, the thermalization rates slow down markedly, as phonon emission begins to compete with impact ionization. Eventually, the electron energy falls below the threshold for ionization, and the electron energy slowly decays to thermal levels via phonon emission, hindered only by phonon absorption.
\begin{figure}
\includegraphics[height=3.25in, width=3.37in]{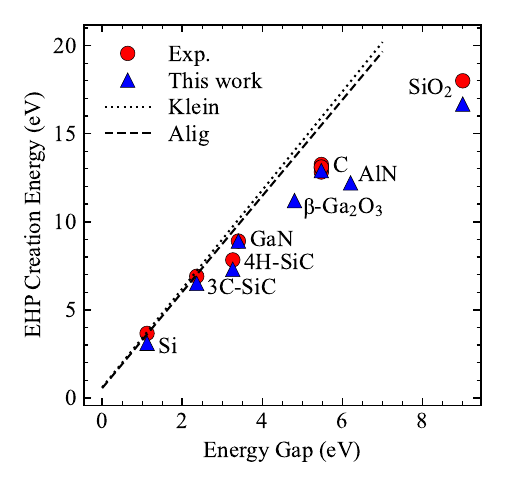}
\caption{\label{fig:creation_en} Measured (red circles) and calculated (blue triangles) EHP creation energies for Si~\cite{scholze2000,lowe2007}, 3C-SiC~\cite{makarov1975a,makarov1975b}, 4H-SiC~\cite{garcia2013,gsponer2024}, GaN~\cite{sellin06}, diamond (C)~\cite{kozlov1975,canali1979,keister2009,morse2007,keister2018}, SiO$_2$~\cite{ausman1975}, $\beta$-Ga$_2$O$_3$, and AlN compared to the analytic expressions of Klein~\cite{klein68} and Alig~{\it et al.}~\cite{alig75}.}
\end{figure}

As the simulation progressed, we tracked the number of pairs generated in the system by each initially ``hot'' electron. Using Eq.~\ref{eq:creation_en}, we then calculated the creation energies of each material. These are shown in Table~\ref{tab:creation_en} and plotted in Fig.~\ref{fig:creation_en}. We note that the magnitude of the creation energy is energy-dependent, as shown by Alig~{\it et al.}~\cite{alig1980}, and that the values shown in Table~\ref{tab:creation_en} were taken for electron kinetic energies at which $\epsilon$ saturates. We will discuss this further, below.
\begin{table}
\caption{Calculated and measured creation energies.}
\label{tab:creation_en}
\setlength{\tabcolsep}{10pt} 
\renewcommand{\arraystretch}{1.2}
\begin{tabular}{c c c}
\toprule
Material&
This work &
Expt. \\
\midrule
AlN & 12.2 & -\\
3C-SiC & 6.6 & 6.90~\cite{makarov1975a,makarov1975b}\\
4H-SiC & 7.3 & 7.83~\cite{gsponer2024}, 7.82~\cite{garcia2013}\\
Si & 3.1 & 3.6~\cite{scholze2000}\\
SiO$_2$ & 16.7 & 18.0~\cite{ausman1975}\\
\bottomrule
\end{tabular}
\end{table}

We observe remarkably good agreement across the board between the measured data available in the literature and our calculated values. One of the largest discrepancies occurs for SiO$_2$ (a difference of $\sim7.6$\%). The measurement of $\epsilon$ for SiO$_2$ was performed in 1975 by Ausman and McClean~\cite{ausman1975}. By fitting a recombination model to current vs. electric field data obtained by Curtis~{\it et al.}~\cite{curtis1974}, Ausman and McClean extracted values of $\epsilon$ for various samples, yielding an average magnitude of $\sim18.0$~eV. However, for at least one of the samples, they extracted a magnitude as low as 16.4~eV, which is in much better agreement with our calculated value of 16.7~eV. Additionally, we note that these measurements were performed in 1975 and, as Alig~{et al.}~\cite{alig1980} remarked, the general trend is that the measured value of $\epsilon$ tends to decrease as new measurements are performed using improved instruments with a higher collection efficiency. The fit, therefore, may improve with modern measurements. That said, there does appear to be a tendency for the calculated value to be slightly lower than its measured counterpart. Most notably, this is the case for Si, for which the magnitude of the creation energy is well established. We will discuss why we think this occurs and our attempt to address it in Sec.~\ref{sec:gwbse_approach}.

\begin{figure}
\includegraphics[height=3.25in, width=3.37in]{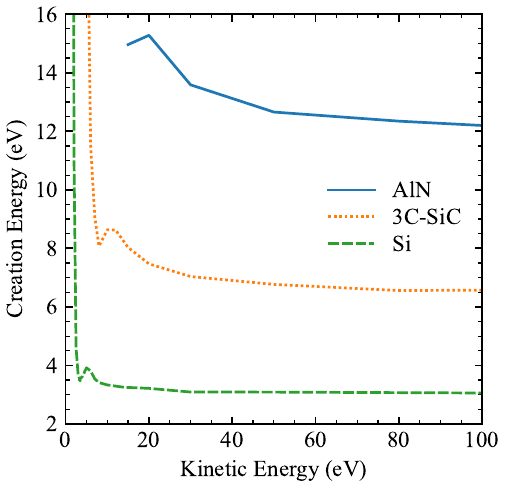}
\caption{\label{fig:creation_en_mult} Calculated creation energies of AlN, 3C-SiC, and Si as a function of electron kinetic energy.}
\end{figure}

Returning to the energy dependence of the creation energy, we have simulated the thermalization of electrons with kinetic energies ranging from 0 to 100~eV in Si, 3C-SiC, and AlN and plotted the results in Fig.~\ref{fig:creation_en_mult}. From the figure, it is clear that $\epsilon$ varies significantly for energies below the plasmon peak and then eventually converges to a constant value. As mentioned above, the magnitudes we reported in Table~\ref{tab:creation_en} were calculated for kinetic energies of $\sim100$~eV, in the energy range in which $\epsilon$ is constant. The creation energy becomes approximately energy-independent as the phonon emission rates become negligible relative to the ELRs (ionization plus plasmon emission) and the ELRs level off to a nearly constant value. Because the saturation of $\epsilon$ occurs at relatively high energies, above the plasmon peak, we conclude that it is important to account for plasmon emission in the simulations. In Sec.~\ref{sec:discussion}, we will discuss further the importance of plasmon emission in calculating $\epsilon$ and why Alig~{\it et al.} obtained reasonable results while ignoring it.

\section{\label{sec:gwbse_approach}The "GW/BSE Approach"}

The GW approach has been shown to improve the calculated {\it ab initio} bandgap for many materials~\cite{shishkin2007,chen2015}, and the BSE has been used to successfully reproduce the experimentally measured dielectric function of Si~\cite{sangalli2016}. In this section, we will show that in using the GW/BSE approach, we were able to improve the agreement between the predicted and measured values of $\epsilon$ for Si. 

In Sec.~\ref{sec:mc_sim}, we showed the results of our calculations of the creation energy for various materials, and we observed that the calculated data tended to consistently fall slightly below the measured values. For example, as shown in Table~\ref{tab:creation_en},  our predicted value of $\epsilon$ for Si was $\sim3.10$~eV, which is about $14.9\%$ lower than the well-known measured value of $\sim3.6$~eV~\cite{scholze2000}. Given the success of the GW approach combined with the BSE in reproducing the experimentally measured dielectric function of Si~\cite{sangalli2016}, we tested its effect on the calculation of the creation energy.

We utilized the Vienna Ab initio Simulation Package~\cite{kresse93,kresse96a,kresse96b} (VASP) to perform the GW calculation of the quasi-particle energy shifts and the dielectric function via the BSE. This began with a standard ground-state DFT calculation, including a relaxation of the crystal structure and a self-consistent solution of the Kohn-Sham equations using a $12\times12\times12$ uniform grid of k-points to converge the Kohn-Sham potential. The relaxed lattice parameter from this step was $\sim5.47$~\AA. With the potential converged, we performed a non-self-consistent calculation of the band structure with 512 unoccupied states (conduction bands). 

The primary difference between the standard Kohn-Sham equations and the GW equations is that the exchange-correlation term is replaced by the so-called self-energy $\Sigma$. In the GW approach, the self-energy is approximated as a Green's function ``G'' multiplied by the screened Coulomb potential ``W''. Each term depends on the wavefunctions, so one should, in principle, employ a self-consistent technique to solve the equations and converge both ``G'' and ``W''. The primary calculation variants or flavors include G$_0$W$_0$, a single-shot variant in which the quasiparticle energies are calculated in a single iteration; GW$_0$, in which the ``W'' is held fixed to the initial DFT value while the ``G'' alone is iterated a user-defined number of times; and GW, in which both terms are iterated. In most cases, GW$_0$ provides the optimal balance of speed and accuracy, and we employ it here.

Finally, we performed a BSE calculation of the dielectric function. The BSE is an equation within the many-body perturbation theory framework used to calculate the excited-state response of materials. Whereas standard DFT calculations employ an independent-particle picture, the BSE accounts for the interaction between the excited electron and hole, known as excitonic effects. As before, we carried out this calculation for several q-points along the first reciprocal lattice vector of Si, again assuming spherical isotropy as in the TDDFT calculations of the loss functions described in Sec.~\ref{sec:abinitio}. The computational cost of each calculation for each q-point renders it challenging to obtain the loss function on a fine mesh. However, the loss functions of the materials studied in this work tend to change relatively little for q-points near $\Gamma$, and due to the $1/q^2$ term in Eq.~\ref{eq:ELR}, these points carry the most weight in the calculation of the ELRs and the determination of final states in the MC simulation. Therefore, relatively few points near $\Gamma$ are required to adequately represent the q-dependence of the loss function with little loss in accuracy.
\begin{figure}
\includegraphics[height=5.0in, width=3.37in]{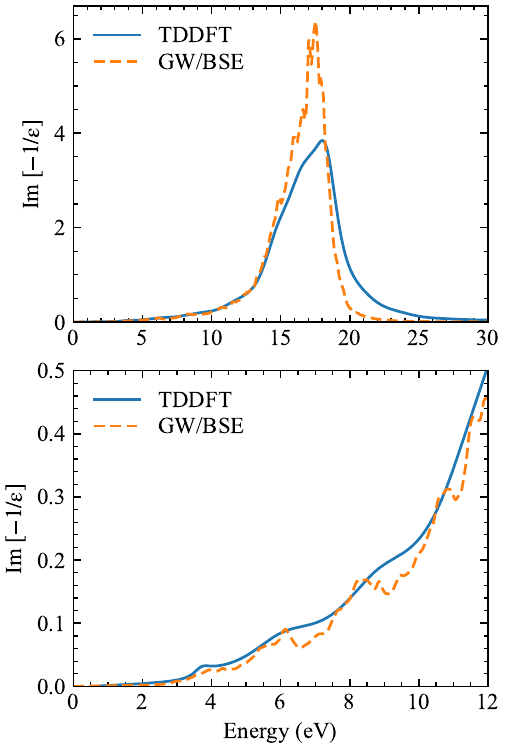}
\caption{\label{fig:lf_si_gw} The loss function of silicon for $q=0.165$~\AA$^{-1}$ calculated using both the TDDFT and GW/BSE approaches. The bottom frame shows the same plot but only from 0 to 12~eV.}
\end{figure}

\begin{figure}
\includegraphics[height=2.5in, width=3.37in]{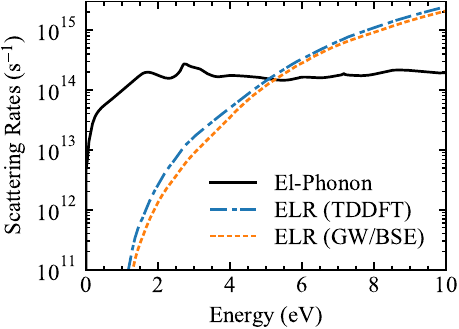}
\caption{\label{fig:scatt_rates_gw} The electron ELRs of Si calculated using the TDDFT and GW/BSE approaches, shown with the electron-phonon scattering rates.}
\end{figure}

In Fig.~\ref{fig:lf_si_gw}, we show the loss function of Si calculated using both the TDDFT and GW/BSE approaches for $q=0.165$~\AA$^{-1}$. Firstly, we note that the plasmon peak in the GW result is narrower and has a larger magnitude. It decays more quickly than the TDDFT curve as the electron energy increases above the plasmon peak, while below the peak, the two curves lie nearly on top of one another. To better observe the differences over the energy range below the plasmon peak, we have included an inset figure for electron energies up to 12~eV. It shows that the GW loss function is slightly lower in magnitude for the majority of the range. Although the extent to which it is lower in magnitude is seemingly small, it visibly affects the ELRs near the intersection with the phonon scattering rates, as shown in Fig.~\ref{fig:scatt_rates_gw}.
\begin{figure}
\includegraphics[height=5.0in, width=3.37in]{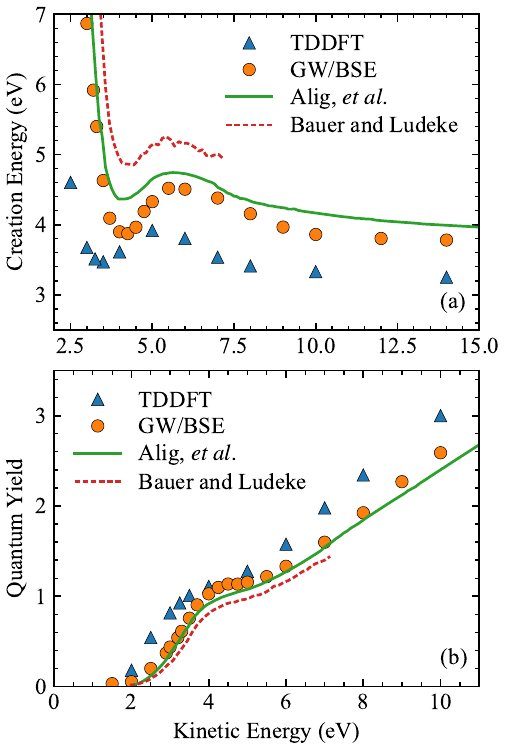}
\caption{\label{fig:qy_creation_en} (a) The creation energy as a function of the kinetic energy of the incident electron. Here, we compare the TDDFT and GW/BSE results with the work of Alig~{\it et al.}~\cite{alig1980} and the measurements of Bauer and Ludeke~\cite{bauer1994}. (b) The quantum yield as a function of kinetic energy.}
\end{figure}

Following the calculation of the loss function and the ELRs, we resimulated the relaxation of electrons in Si and obtained a value of $\epsilon$ of $\sim3.50$~eV. This result is a significant improvement in agreement with the measured value of $\sim3.6$~eV. We also calculated the quantum yield and the dependence of $\epsilon$ on the incident electron kinetic energy. These are shown in Fig.~\ref{fig:qy_creation_en} with those calculated by Alig~{\it et al.} and those measured by Bauer and Ludeke~\cite{bauer1994}. In all cases, the use of the GW/BSE loss function markedly improved the accuracy of the predicted results. We also note the relatively large effect of seemingly minute changes in the loss function in Fig.~\ref{fig:lf_si_gw}. From this, it is clear that the rate ratio near the intersection of the ELRs and phonon scattering rates is largely responsible for determining the value of $\epsilon$. It can be expected then that further improvements in the prediction will likely come from relatively small changes in the way the phonon scattering rates and ELRs are calculated (i.e., denser {\bf  k} and {\bf q} Brillouin zone meshes in the calculations of the band structure, phonon dispersion, electron-phonon matrix elements, and dielectric function, variations in the pseudopotentials used, the method employed in handling the exchange-correlation term, etc.).

\section{\label{sec:discussion} Discussion}

\subsection{\label{sec:q1} Is it accurate to approximate $A$ as a constant?}

Given the success of the methods utilized by Alig~{\it et al.}~\cite{alig1980,alig1983}, for many materials, it is perhaps appropriate to wonder whether there is something fundamental about the assumption that $A$ is constant. Indeed, a constant value of $A$ suggests that the ratio of the phonon scattering matrix elements to those of ionization is constant, meaning that the details of the momentum- and energy-dependence of the matrix elements and those of the band structure are negligible for this calculation.

To begin to answer this question, we first rearranged Eq.~\ref{eq:rate_ratio} as follows:
\begin{equation}
\label{eq:rate_ratio2}
A=\frac{r^{({\rm ph})}(E)}{r^{({\rm ii})}(E)}\frac{2\pi}{105}\frac{(E-E_{\rm g})^{7/2}}{(E-\hbar\omega_0)^{1/2}}.
\end{equation}
For $\hbar\omega_0$, we substituted the longitudinal optical phonon energy (0.063~eV for Si), for $r^{({\rm ph})}(E)$, we used the carrier-phonon scattering rates averaged over equi-energy shells (shown in Fig.~\ref{fig:scatt_rates}), and similarly for $r^{({\rm ii})}(E)$, we used the ELRs averaged over equi-energy shells.

\begin{figure}
\includegraphics[height=4.0in, width=3.37in]{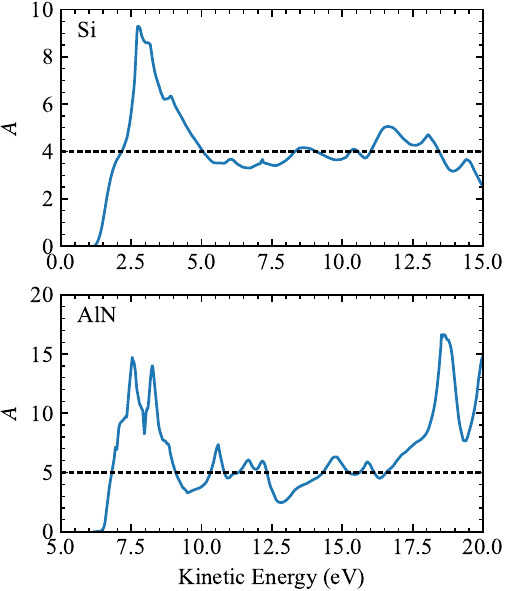}
\caption{\label{fig:A_si_aln} The parameter $A$ of Si and AlN, calculated using Eq.~\ref{eq:rate_ratio2}. For Si, the GW/BSE ELRs were used. The dashed horizontal lines indicate possible constant values of $A$ that could be used in the methods of Alig~{\it et al.}~\cite{alig1980}.}
\end{figure}

In Fig.~\ref{fig:A_si_aln}, we show $A$ as a function of energy for Si and AlN. For Si, we used the ELRs calculated with the GW/BSE approach to calculate $A$. Below $\sim5$~eV, for Si, $A$ is highly variable, while between $\sim5$-13~eV, the magnitude of $A$ flattens considerably, hovering around $\sim4$. This is relatively close to the value of 5.2 used by Alig~{\it et al.}. One can therefore say that near the intersection (at $\sim5.24$~eV) of the phonon scattering and ionization rates (where the competition between these processes is most important), $A$ is approximately constant. This fact partly explains why this method works so well for Si. Similarly, for AlN, we observe a flattening of the parameter $A$ near the intersection of the rates, with a magnitude hovering around a constant value of $\sim5$.

\begin{figure}
\includegraphics[height=6.0in, width=3.37in]{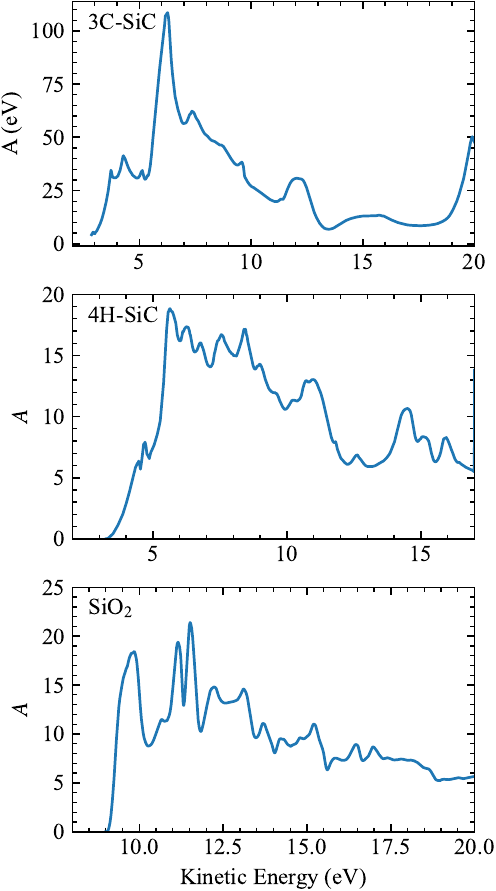}
\caption{\label{fig:A_sic_sio2} The parameter $A$ of 3C-SiC, 4H-SiC and SiO$_2$, calculated using Eq.~\ref{eq:rate_ratio2}.}
\end{figure}

Conversely, in Fig.~\ref{fig:A_sic_sio2}, we also show that for both polymorphs of SiC and for SiO$_2$, $A$ is not remotely constant over the relevant energy range (near the rate intersection). We can therefore conclude that, while $A$ may be approximated as constant for some materials, the assumption that it is constant and indeed identical for all semiconductors and dielectrics is unfounded. 


\subsection{\label{sec:q2} Can plasmon emission be ignored?}

One of the conclusions of the paper by Alig~{\it et al.} is that plasmon emission can be ignored when calculating the creation energy. Given that Alig~{\it et al.} showed that the creation energy varies with energy and eventually converges to a constant value well above the plasmon energy, this conclusion seems to contradict Pine's~\cite{pines56} claim that plasmon emission is the dominant energy-loss process in the 10-100~eV range. And the success of the present work (which accounts for plasmon emission) in predicting accurate values of $\epsilon$ also suggests that plasmon emission is an important part of the calculation. Why then were Alig~{\it et al.} successful in calculating $\epsilon$ without plasmon emission, and why, when they included it, did their predictions worsen?

Part of the answer lies in the fact that the parameter $A$ was specifically fitted to 5.2 so that the method excluding plasmon emission would succeed. Likely, fitting $A$ separately with plasmon emission included would yield better agreement than simply using the same value of 5.2 (fitted for the case without it). The other part of the answer lies in the ionization rate. Alig~{\it et al.} calculated the ionization rate using Kane's random-k approximation. In their approach, however, they employed free-electron bands, while we have employed the full band structure calculated using DFT. The density of states for free electrons is different, and the ionization rates via the random-k approximation are strongly dependent on the density of states, so one may expect significant differences between the ionization rates calculated using free-electron bands and those calculated using full bands.

\begin{figure}
\includegraphics[height=3.25in, width=3.37in]{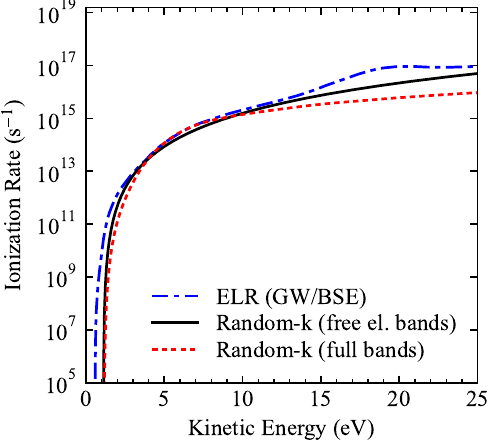}
\caption{\label{fig:free_el_ii} A plot of the ionization rates for free-electron bands within the random-k approximation for Si. Also included are random-k ionization rates calculated using the full DFT band structure and the ELRs calculated using the GW/BSE dielectric function.}
\end{figure}

In Fig.~\ref{fig:free_el_ii}, we show the random-k ionization rates calculated using free-electron bands and those calculated using the full band structure of Si. Additionally, we show the ELRs of Si, calculated using the GW/BSE dielectric function. Interestingly, the free-electron ionization rates resemble the ELRs closely but are markedly different from the full-band random-k rates, especially as the kinetic energy increases. This suggests that by employing free-electron bands, one, perhaps by accident, nearly exactly accounts for both impact ionization and plasmon emission without additional effort. Therefore, the work of Alig~{\it et al.} essentially includes plasmon emission by default, which helps to explain why it is so successful in predicting the energy dependence of the quantum yield and $\epsilon$ for Si.

\subsection{\label{sec:q3} The quantum yield of Si}

Lastly, in this section, we briefly discuss the shape of the quantum yield curve of Si, as shown in Fig.~\ref{fig:qy_creation_en}(b). Specifically, of interest is the ``bump'' that occurs near $\sim4$~eV, which appears in the calculated results of this work and that of Alig~{\it et al.}~\cite{alig1980} as well as the measured data of Bauer and Ludeke~\cite{bauer1994}. Above $\sim5.5$~eV, the curve becomes approximately linear and no further ``bumps'' are observed. Here, we will explain the origin of this feature and why the curve becomes smooth as the kinetic energy increases.

The fact that we observe this ``bump'' in the results of Alig~{\it et al.} suggests that its origin is not related to the band structure or the (joint) density of states of the material, as Alig~{\it et al.} employed free-electron bands. Additionally, one may immediately reject the idea that the scattering matrix elements play a role, due to the use of the ``scattering-rate'' assumption. Instead, one must consider the ``probability method'' of Alig~{\it et al.} as a possible origin.

\begin{figure}
\includegraphics[height=2.5in, width=3.37in]{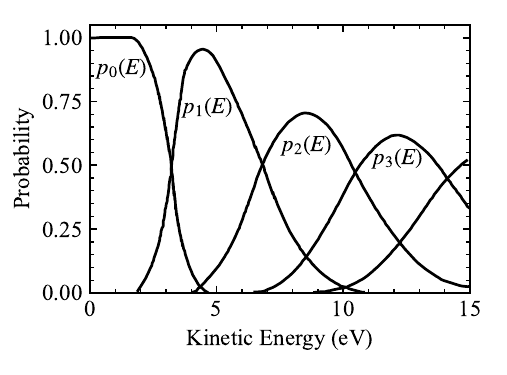}
\caption{\label{fig:prob_n} Curves representing $p_n(E)$ for a few values of $n$ as calculated by Alig~{et al.}~\cite{alig1980}.}
\end{figure}

As described above, the ``probability method'' involves calculating the probabilities $p_n(E)$ that a carrier of kinetic energy $E$ generates exactly $n$ pairs. In Fig.~\ref{fig:prob_n}, we show the results of this calculation as published by Alig~{\it et al.}~\cite{alig1980}. Between 4 and 5~eV, $p_1(E)$ peaks at a magnitude of $\sim0.95$, meaning nearly all carriers with kinetic energies in this range generate exactly 1 pair. In Fig.~\ref{fig:qy_creation_en}(b), the ``bump'' in the quantum yield occurs in this same energy range and the magnitude if the quantum yield at this ``bump'' is $\sim1$. Therefore, the correct interpretation is that the quantum yield flattens at $\sim1$ in the energy range of approximately 4-5~eV, as nearly all carriers excite exactly 1 pair. As the energy decreases, the average number of pairs generated goes below one, as $p_0(E)$ increases.

Above $\sim5.5$~ eV, the quantum yield curve is smooth, resulting from the fact that each successive $p_n(E)$ curve is broader and overlaps more with the previous one.

\section{\label{sec:conclusions} Conclusions}

Employing a first-principles approach, we have calculated the band structure, phonon dispersion, dielectric function, and electron-phonon matrix elements for AlN, 3C-SiC, 4H-SiC, Si, and SiO$_2$. Using these material properties, we have evaluated the carrier-phonon scattering rates and the energy-loss rates, which we have incorporated into a full-band MC simulation of the relaxation and transport of electrons and holes. By simulating the thermalization of electrons with kinetic energies as high as 100~eV, we have calculated the creation energies of the listed materials (AlN: 12.2~eV, 3C-SiC: 6.5~eV, 4H-SiC: 7.3~eV, Si: 3.1~eV, and SiO$_2$: 16.7~eV) as well as their energy dependence and that of the quantum yield.

We have shown that this approach yields results that are in good agreement with measured data. We have also shown that improvements in the agreement, in the case of Si, can be made by employing the GW approach to calculate the band structure and the BSE to calculate the dielectric function. Whereas the TDDFT approach yielded a value of $\epsilon\sim3.10$~eV, the GW/BSE approach yielded a value of $\sim3.5$~eV, which is in much better accord with the experimental value of $\sim3.6$~eV. Similar improvements have also been shown in the energy dependence of the quantum yield.  

We have also discussed the parameter $A$ from the work of Alig~{\it et al.}~\cite{alig1980} and shown that while it may be approximated as constant for Si and AlN over a relatively wide energy range near the intersection of the phonon-scattering and ionization rates, this is not the case for 3C-SiC, 4H-SiC, or SiO$_2$. Additionally, we showed that plasmon emission must be considered in the calculation of the creation energy, as the creation energy reaches a constant value well above the plasmon peak energy. And finally, we discussed the origin of the ``bump'' in the quantum yield of Si that occurs near a kinetic energy of $\sim4$~eV, and we found that it is related to the probability of generating exactly one EHP.

\section{\label{sec:conclusions} Acknowledgements}

We would like to acknowledge the the computational support of the Texas Advanced Computing Center (TACC) at the University of Texas at Austin for providing high-performance computing. This work has been supported by the Air Force Office of Scientific Research: Award No. FA9550-23-1-0549.

\bibliography{ehp_gw_paper}

@PREAMBLE{
 "\providecommand{\noopsort}[1]{}" 
 # "\providecommand{\singleletter}[1]{#1}%" 
}

@ARTICLE{nielsen23a,
   title        = {{First-principles approach to closing the 10-100 eV gap for charge-carrier thermalization in semiconductors}},
   author       = {Nielsen, D. O. and Van de Walle, C. G. and Pantelides, S. T. and Schrimpf, R. D. and Fleetwood, D. M. and Fischetti, M. V.},
   year         = {2023},
   journal      = {Phys. Rev. B},
   volume       = {108},
   number = {15},
   month = {10},
   pages        = {155203},
   usera = {Art. no.},
   doi          = {https://doi.org/10.1103/PhysRevB.108.155203},
}

@ARTICLE{nielsen23b,
   title        = {{Thermalization of radiation-induced electrons in wide-bandgap materials: A first-principles approach}},
   author       = {Nielsen, D. O. and Fischetti, M. V.},
   year         = {2023},
   journal      = {Appl. Phys. Lett.},
   volume       = {123},
   number = {25},
   month = {12},
   pages        = {252107},
   usera = {Art. no.},
   doi          = {https://doi.org/10.1063/5.0186802},
}

@ARTICLE{nielsen26a,
  author={Nielsen, D. O. and Pantelides, S. T. and Schrimpf, R. D. and Fleetwood, D. M. and Fischetti, M. V.},
  journal={IEEE Trans. Nucl. Sci.}, 
  title={First-Principles Study of Radiation-Induced Electron-Hole Pair Generation and Transport in {GaN} and {Al$_{0.25}$Ga$_{0.75}$N/GaN HEMTs}}, 
  year={2026},
  volume={73},
  number={4},
  pages={1355-1369},
  doi={10.1109/TNS.2026.3669001}
  }

@ARTICLE{shockley61,
   title        = {Problems related to p-n junctions in silicon},
   author       = "W. Shockley",
   year         = "1961",
   journal      ={Solid-State Electron.},
   volume       = {2},
   number = {1},
   pages        = {35-67},
   month={02},
   doi          = {https://doi.org/10.1016/0038-1101(61)90054-5},
}

@ARTICLE{klein68,
   title        = {Bandgap dependence and related features of radiation ionization energies in semiconductors},
   author       = {Klein, C. A.},
   year         = "1968",
   journal    = "J. Appl. Phys.",
   volume    = "39",
   number = {4},
   pages     = {2029-2038},
   month     = {03},
   doi          = {https://doi.org/10.1063/1.1656484},
}

@ARTICLE{kresse93,
   title        = {Ab initio molecular dynamics for liquid metals},
   author       = {Kresse, G. and Hafner, J.},
   year         = "1993",
   journal      = "Phys. Rev. B",
   volume       = "47",
   pages        = "558",
   doi          = {https://doi.org/10.1103/PhysRevB.47.558},
}

@ARTICLE{kresse96a,
   title        = {Efficiency of ab-initio total energy calculations for metals and semiconductors using a plane-wave basis set},
   author       = {Kresse, G. and Furthm{\"u}ller, J.},
   year         = "1996",
   journal      = "Comput. Mat. Sci.",
   volume       = "6",
   pages        = "15",
   doi          = {https://doi.org/10.1016/0927-0256(96)00008-0},
}

@ARTICLE{kresse96b,
   title        = {Efficient iterative schemes for ab initio total-energy calculations using a plane-wave basis set},
   author       = {Kresse, G. and Furthm{\"u}ller, J.},
   year         = "1996",
   journal      = "Phys. Rev. B",
   volume       = "54",
   pages        = "11169",
   doi          = {https://doi.org/10.1103/PhysRevB.54.11169},
}

@ARTICLE{schlipf15,
   title        = {{Optimization algorithm for the generation of ONCV pseudopotentials}},
   author       = {Schlipf, M. and Gygi, F.},
   journal      = "Comput. Phys. Commun.",
   year         = "2015",
   volume       = "196",
   pages        = "36",
   doi          = {https://doi.org/10.1016/j.cpc.2015.05.011},
}

@article{perdew96,
  title={{Generalized gradient approximation made simple}},
  author={Perdew, J. P. and Burke, K. and Ernzerhof, M.},
  journal={Phys. Rev. Lett.},
  year={1996},
  volume={77},
  pages={3865-3868},
  doi={https://doi.org/10.1103/PhysRevLett.77.3865}
}

@article{ponce16,
  title={{EPW: Electron-phonon coupling, transport and superconducting properties using maximally localized Wannier functions}},
  author={Poncé, S. and Margine, E. R. and Verdi, C. and Giustino, F.},
  journal={Comput. Phys. Commun.},
  year={2016},
  volume={209},
  pages={116-133},
  month={12},
  doi={https://doi.org/10.1016/j.cpc.2016.07.028}
}

@article{giustino07,
  title={{Electron-phonon interaction using Wannier functions}},
  author={Giustino, F. and Cohen, M. L. and Louie, S. G.},
  journal={Phys. Rev. B},
  year={2007},
  volume={76},
  month={10},
  number = {16},
  pages={165108},
  usera={Art. no.},
  doi={https://doi.org/10.1103/PhysRevB.76.165108}}

@article{pines56,
  title={Collective Energy Losses in Solids},
  author={Pines, D.},
  journal={Rev. Mod. Phys.},
  month={07},
  year={1956},
  volume={28},
  pages={184-198},
  number = {3},
  doi={https://doi.org/10.1103/RevModPhys.28.184}
}

@article{quinn58,
  title={Electron Self-Energy Approach to Correlation in a Degenerate Electron Gas},
  author={Quinn, J. J. and Ferrell, R. A.},
  journal={Phys. Rev.},
  month={11},
  year={1958},
  volume={112},
  pages={812-827},
  number = {3},
  doi={https://doi.org/10.1103/PhysRev.112.812}
}

@article{quinn62,
  title={Range of Excited Electrons in Metals},
  author={Quinn, J. J.},
  journal={Phys. Rev.},
  year={1962},
  volume={126},
  pages={1453-1457},
  number={4},
  month={05},
  doi={https://doi.org/10.1103/PhysRev.126.1453}
}

@article{penn87,
  title={{Electron mean-free-path calculations using a model dielectric function}},
  author={Penn, D. R.},
  journal={Phys. Rev. B},
  year={1987},
  volume={35},
  number = {2},
  month = {01},
  pages={482-486},
  doi={https://doi.org/10.1103/PhysRevB.35.482}
}

@article{giannozzi09,
  title={{QUANTUM ESPRESSO: a modular and open-source software project for quantum simulations of materials}},
  author={Giannozzi, Paolo and Baroni, Stefano and Bonini, Nicola and Calandra, Matteo and Car, Roberto and Cavazzoni, Carlo and Ceresoli, Davide and 
          Chiarotti, Guido L and Cococcioni, Matteo and Dabo, Ismaila and Dal Corso, Aandrea and de Gironcoli, Stefano and Fabris, Stefano and 
          Fratesi, Guido and Gebauer, Ralph and Gerstmann, Uwe and Gougoussis, Christos and Kokalj, Anton and Lazzeri, Michele and Martin-Samos, Layla 
          and Marzari, Nicola and Mauri, Francesco and Mazzarello, Riccardo and Paolini, Stefano and Pasquarello, Alfredo and Paulatto, Lorenzo 
          and Sbraccia, Carlo and Scandolo, Sandro and Sclauzero, Gabriele and Seitsonen, Ari P. and Smogunov, Alexander and Umari, Paolo and 
          Wentzcovitch, Renata M.},
  journal={J. Phys.: Cond. Matt.},
  volume={21},
  number={39},
  month = {09},
  pages={395502},
  usera = {Art. no.},
  year={2009},
  publisher={IOP Publishing},
  doi={https://doi.org/10.1088/0953-8984/21/39/395502}
}

@article{timrov15,
  title={{turboEELS - A code for the simulation of the electron energy loss and inelastic X-ray scattering spectra using the Liouville-Lanczos approach to time-dependent density-functional perturbation theory}},
  author={Timrov, I. and Vast, N. and Gebauer, R. and Baroni, S.},
  journal={Comput. Phys. Commun.},
  month = {11},
  year={2015},
  volume={196},
  pages={460-469},
  doi={https://doi.org/10.1016/j.cpc.2015.05.021}
}

@article{bethe30,
author = {Bethe, H.},
title = {Zur Theorie des Durchgangs schneller Korpuskularstrahlen durch Materie},
journal = {Annalen der Physik},
volume = {397},
number = {3},
pages = {325-400},
doi = {https://doi.org/10.1002/andp.19303970303},
year = {1930}
}

@article{sellin06,
title = {New materials for radiation hard semiconductor dectectors},
journal = {Nucl. Instrum. Methods Phys. Res. A},
volume = {557},
number = {2},
pages = {479-489},
month = {02},
year = {2006},
doi = {https://doi.org/10.1016/j.nima.2005.10.128},
author = {P.J. Sellin and J. Vaitkus},
}

@article{jacoboni83,
  title={{The Monte Carlo method for the solution of charge transport in semiconductors with applications to covalent materials}},
  author={Jacoboni, C. and Reggiani, L.},
  journal={Rev. Mod. Phys.},
  year={1983},
  number = {3},
  volume={55},
  pages={645-705},
  month = {07},
  doi={https://doi.org/10.1103/RevModPhys.55.645}
}

@article{fischetti95,
    author = {Fischetti, M. V. and Laux, S. E. and Crabb\'e, E.},
    title = {Understanding hot‐electron transport in silicon devices: {Is} there a shortcut?},
    journal = {J. Appl. Phys.},
    volume = {78},
    number = {2},
    pages = {1058-1087},
    year = {1995},
    month={07},
    doi = {10.1063/1.360340}
}

@article{fang19,
  title = {Electron Transport Properties of ${\mathrm{Al}}_{x}{\mathrm{Ga}}_{1\ensuremath{-}x}\mathrm{N}/\mathrm{Ga}\mathrm{N}$ Transistors Based on First-Principles Calculations and {Boltzmann}-Equation {Monte Carlo} Simulations},
  author = {Fang, Jingtian and Fischetti, Massimo V. and Schrimpf, Ronald D. and Reed, Robert A. and Bellotti, Enrico and Pantelides, Sokrates T.},
  journal = {Phys. Rev. Appl.},
  volume = {11},
  issue = {4},
  pages = {044045},
  usera={Art. no.},
  numpages = {14},
  year = {2019},
  month = {4},
  doi = {10.1103/PhysRevApplied.11.044045}
}

@article{blochl94,
  title={{Improved tetrahedron method for Brillouin-zone integrations}},
  author={Bl{\"o}chl, P. E. and Jepsen, O. and Andersen, O. K.},
  journal={Phys. Rev. B},
  year={1994},
  volume={49 23},
  pages={16223-16233},
  doi={https://doi.org/10.1103/PhysRevB.49.16223}
}

@article{lappe61,
title = {The energy of electron-hole pair formation by {X-rays in PbO}},
journal = {J. Phys. Chem. Solids},
volume = {20},
number = {3},
pages = {173-176},
year = {1961},
month={08},
doi = {https://doi.org/10.1016/0022-3697(61)90001-4},
author = {F. Lappe}
}

@article{mckay1951,
  title = {Electron-Hole Production in Germanium by Alpha-Particles},
  author = {McKay, K. G.},
  journal = {Phys. Rev.},
  volume = {84},
  issue = {4},
  pages = {829-832},
  year = {1951},
  month = {Nov},
  publisher = {American Physical Society},
  doi = {10.1103/PhysRev.84.829},
}

@article{alig75,
  title = {Electron-Hole-Pair Creation Energies in Semiconductors},
  author = {Alig, R. C. and Bloom, S.},
  journal = {Phys. Rev. Lett.},
  volume = {35},
  issue = {22},
  pages = {1522--1525},
  year = {1975},
  month = {Dec},
  publisher = {American Physical Society},
  doi = {10.1103/PhysRevLett.35.1522},
}

@article{ahearn48,
  title = {Conductivity Induced in Diamond by Alpha-Particle Bombardment and Its Variation among Specimens},
  author = {Ahearn, A. J.},
  journal = {Phys. Rev.},
  volume = {73},
  issue = {9},
  pages = {1113},
  year = {1948},
  month = {May},
  publisher = {American Physical Society},
  doi = {10.1103/PhysRev.73.1113},
  url = {https://link.aps.org/doi/10.1103/PhysRev.73.1113}
}

@article{mckay1953,
  title = {Electron Multiplication in Silicon and Germanium},
  author = {McKay, K. G. and McAfee, K. B.},
  journal = {Phys. Rev.},
  volume = {91},
  issue = {5},
  pages = {1079-1084},
  year = {1953},
  month = {Sep},
  publisher = {American Physical Society},
  doi = {10.1103/PhysRev.91.1079},
}

@article{baldinger1961,
title = {On the energy expended per electron-hole-pair produced in p-n-junction-detectors},
journal = {Nucl. Instrum. Methods},
volume = {10},
pages = {237-239},
year = {1961},
month = {12},
doi = {https://doi.org/10.1016/S0029-554X(61)80108-0},
author = {Baldinger, E. and Czaja, W.}
}

@article{alig1980,
  title = {Scattering by ionization and phonon emission in semiconductors},
  author = {Alig, R. C. and Bloom, S. and Struck, C. W.},
  journal = {Phys. Rev. B},
  volume = {22},
  issue = {12},
  pages = {5565--5582},
  year = {1980},
  month = {12},
  publisher = {American Physical Society},
  doi = {10.1103/PhysRevB.22.5565},
}

@article{alig1983,
  title = {Scattering by ionization and phonon emission in semiconductors. {II. Monte Carlo} calculations},
  author = {Alig, R. C.},
  journal = {Phys. Rev. B},
  volume = {27},
  issue = {2},
  pages = {968--977},
  numpages = {0},
  year = {1983},
  month = {01},
  doi = {10.1103/PhysRevB.27.968},
}

@article{roosbroeck1965,
  title = {Theory of the Yield and {Fano} Factor of Electron-Hole Pairs Generated in Semiconductors by High-Energy Particles},
  author = {van Roosbroeck, W.},
  journal = {Phys. Rev.},
  volume = {139},
  issue = {5A},
  pages = {A1702--A1716},
  year = {1965},
  month = {08},
  publisher = {American Physical Society},
  doi = {https://doi.org/10.1103/PhysRev.139.A1702},
}

@article{antoncik1968,
  title = {On the theory of quantum efficiency in germanium},
  author = {Antoncik, E.},
  journal = {Czech. J. Phys. B},
  volume = {18},
  issue = {2},
  pages = {157-170},
  year = {1968},
  month = {02},
  doi = {10.1007/BF02101443},
}

@article{drummond1971,
    author = {Drummond, W. E. and Moll, J. L.},
    title = {Hot Carriers in {Si and Ge} Radiation Detectors},
    journal = {J. Appl. Phys.},
    volume = {42},
    number = {13},
    pages = {5556-5562},
    year = {1971},
    month = {12},
    doi = {10.1063/1.1659981},
}

@article{fischetti88,
  title = {{Monte Carlo analysis of electron transport in small semiconductor devices including band-structure and space-charge effects}},
  author = {Fischetti, M. V. and Laux, S. E.},
  journal = {Phys. Rev. B},
  volume = {38},
  issue = {14},
  pages = {9721--9745},
  year = {1988},
  month = {11},
  publisher = {American Physical Society},
  doi = {https://doi.org/10.1103/PhysRevB.38.9721},
}

@article{tang1983,
    author = {Tang, J. Y. and Hess, K.},
    title = {Impact ionization of electrons in silicon (steady state)},
    journal = {J. Appl. Phys.},
    volume = {54},
    number = {9},
    pages = {5139-5144},
    year = {1983},
    month = {09},
    doi = {10.1063/1.332737},
}

@article{shichijo1981,
  title = {Band-structure-dependent transport and impact ionization in {GaAs}},
  author = {Shichijo, H. and Hess, K.},
  journal = {Phys. Rev. B},
  volume = {23},
  issue = {8},
  pages = {4197--4207},
  year = {1981},
  month = {04},
  doi = {10.1103/PhysRevB.23.4197},
}

@article{rothwarf1973,
    author = {Rothwarf, A.},
    title = {{Plasmon theory of electron‐hole pair production: efficiency of cathode ray phosphors}},
    journal = {J. Appl. Phys.},
    volume = {44},
    number = {2},
    pages = {752-756},
    year = {1973},
    doi = {10.1063/1.1662257}
}

@article{ausman1975,
    author = {Ausman, Jr., G. A. and McLean, F. B.},
    title = {{Electron-hole pair creation energy in SiO$_2$}},
    journal = {Appl. Phys. Lett.},
    volume = {26},
    number = {4},
    pages = {173-175},
    year = {1975},
    doi = {10.1063/1.88104}
}

@article{kingsley1970,
doi = {10.1149/1.2407510},
year = {1970},
volume = {117},
number = {3},
pages = {353},
author = {Kingsley, J. D. and Ludwig, G. W.},
title = {{The efficiency of cathode‐ray phosphors: II . Correlation with other properties}},
journal = {J. Electrochem. Soc.}
}

@article{lioliou2022,
  title = {Electron-hole pair creation and conversion efficiency in radioisotope microbatteries},
  journal = {Appl. Radiat. Isot.}, 
  volume = {180},
  pages = {110042},
  year = {2022},
  month = {02},
  doi = {https://doi.org/10.1016/j.apradiso.2021.110042},
  author = {G. Lioliou and A. M. Barnett},
  }

@article{whitaker2020,
   doi = {10.1088/1361-6641/ab9f8e},
   year = {2020},
   month = {08},
   volume = {35},
   number = {9},
   pages = {095026},
   author = {Whitaker, M. D. C. and Lioliou, G. and Krysa, A. B. and Barnett, A. M.},
   title = {Al$_{0.6}$Ga$_{0.4}$As x-ray avalanche photodiodes for spectroscopy},
   journal = {Semiconductor Science and Technology},
   }

@article{curtis1974,
    author = {Curtis, O. L., Jr. and Srour, J. R. and Chiu, K. Y.},
    title = {Hole and electron transport in {SiO$_2$} films},
    journal = {J. Appl. Phys.},
    volume = {45},
    number = {10},
    pages = {4506-4513},
    year = {1974},
    month = {10},
    doi = {10.1063/1.1663079}
    }

@article{dixon2016,
   author = {Dixon, J. and Rajan, A. and Bohlemann, S. and Coso, D. and Upadhyaya, A. D. and Rohatgi, A. and Chu, S. and Majumdar, A. and Yee, S.},
   doi = {10.1038/srep38182},
   year = {2016},
   month = {12},
   volume = {6},
   number = {1},
   pages = {38182},
   title = {Evaluation of a Silicon {$^90$Sr} Betavoltaic Power Source},
   journal = {Sci. Rep.},
   }

@article{olsen2012,
   author = {Olsen, L. C. and Cabauy, P. and Elkind, B. J.},
   doi = {10.1063/PT.3.1820},
   year = {2012},
   month = {12},
   volume = {65},
   pages = {35},
   title = {Betavoltaic power sources},
   journal = {Phys. Today},
   }

@article{shrestha2017,
   author = {Shrestha, S. and
Fischer, R. and
Matt, G. J. and
Feldner, P. and
Michel, T. and
Osvet, A. and
Levchuk, I. and
Merle, B. and
Golkar, S. and
Chen, H. and
Tedde, S. F. and
Schmidt, O. and
Hock, R. and
Rührig, M. and
Göken, M. and
Heiss, W. and
Anton, G. and
Brabec, C J.},
   doi = {10.1038/nphoton.2017.94},
   year = {2017},
   month = {07},
   volume = {6},
   number = {1},
   pages = {436-440},
   title = {High-performance direct conversion {X}-ray detectors based on sintered hybrid lead triiodide perovskite wafers},
   journal = {Nat. Photonics},
   }

@article{liu2020,
   author = {Liu, J. and Xia, F. and Xiao, D. and Garc\'ia de Abajo and F. J. and Sun, D.},
   doi = {10.1038/nphoton.2017.94},
   year = {2020},
   month = {08},
   volume = {19},
   number = {8},
   pages = {830-837},
   title = {Semimetals for high-performance photodetection},
   journal = {Nat. Mater.},
   }

@article{fano1947,
   author = {Fano, U.},
   doi = {https://doi.org/10.1103/PhysRev.72.26},
   year = {1947},
   month = {07},
   volume = {72},
   number = {1},
   pages = {26-29},
   title = {Ionization Yield of Radiations. {II. The} Fluctuations of the Number of Ions},
   journal = {Phys. Rev.},
   }

@article{david2016,
   author = {David, J.},
   title = {The staircase photodiode},
   journal = {Nat. Photonics},
   doi = {https://doi.org/10.1038/nphoton.2016.98},
   year = {2016},
   month = {06},
   volume = {10},
   number = {6},
   pages = {364-366},
   }

@article{polman2016,
   author = {Polman, A. and Knight, M. and Garnett, E. C. and Ehrler, B. and Sinke, W. C.},
   title = {Photovoltaic materials: {P}resent efficiencies and future challenges},
   journal = {Science},
   volume = {352},
   number = {6283},
   pages = {aad4424},
   year = {2016},
   doi = {10.1126/science.aad4424}
   }

@article{arp2019,
   author = {Arp, T. B. and Pleskot, D. and Aji, V. and Gabor, N. M.},
   title = {Electron–hole liquid in a van der {Waals} heterostructure photocell at room temperature},
   journal = {Nat. Photonics},
   volume = {13},
   number = {4},
   pages = {245-250},
   year = {2019},
   doi = {10.1038/s41566-019-0349-y}
   }

@article{makarov1975a,
   author = {Makarov, V. V.},
   title = {Energy of electron-hole pair formation along tracks of light particles},
   journal = {Sov. Phys. - Semicond.},
   volume = {9},
   number = {},
   pages = {526},
   year = {1975},
   doi = {}
   }

@article{makarov1975b,
   author = {Makarov, V. V.},
   title = {Spatial distribution of the deceleration of electrons in solids},
   journal = {Sov. Phys. - Semicond.},
   volume = {9},
   number = {},
   pages = {722},
   year = {1975},
   doi = {}
   }

@article{gsponer2024,
 title = {Measurement of the electron–hole pair creation energy in a {4H-SiC} p-n diode},
 journal = {Nuc. Instrum. Methods Phys. Res. A},
 volume = {1064},
 pages = {169412},
 year = {2024},
 doi = {https://doi.org/10.1016/j.nima.2024.169412},
 author = {Andreas Gsponer and Matthias Knopf and Philipp Gaggl and Jürgen Burin and Simon Waid and Thomas Bergauer},
 month={07},
 number={}
 }

@article{garcia2013,
    author = {Garcia, T. R. and Kumar, A. and Reinke, B. and Blue, T. E. and Windl, W.},
    title = {Electron-hole pair generation in {SiC} high-temperature alpha particle detectors},
    journal = {Appl. Phys. Lett.},
    volume = {103},
    number = {15},
    pages = {152108},
    year = {2013},
    month = {10},
    doi = {10.1063/1.4824774}}

@ARTICLE{kozlov1975,
  author={Kozlov, S. F. and Stuck, R. and Hage-Ali, M. and Siffert, P.},
  journal={IEEE Trans. Nucl. Sci.}, 
  title={Preparation and Characteristics of Natural Diamond Nuclear Radiation Detectors}, 
  year={1975},
  volume={22},
  number={1},
  pages={160-170},
  doi={10.1109/TNS.1975.4327634}}

@article{canali1979,
  title = {Electrical properties and performances of natural diamond nuclear radiation detectors},
  journal = {Nucl. Instrum. Methods},
  volume = {160},
  number = {1},
  pages = {73-77},
  year = {1979},
  doi = {https://doi.org/10.1016/0029-554X(79)90167-8},
  author = {Canali, C. and Gatti, E. and Kozlov, S. F. and Manfredi, P. F. and Manfredotti, C. and Nava, F. and Quirini, A.}}

@article{keister2009,
  title = {{Single crystal diamond photodiode for soft X-ray radiometry}},
  journal = {Nucl. Instrum. Methods Phys. Res. A},
  volume = {606},
  number = {3},
  pages = {774-779},
  year = {2009},
  doi = {https://doi.org/10.1016/j.nima.2009.04.044},
  author = {Keister, J. W. and Smedley, J.}}

@article{morse2007, 
  title={Single Crystal {CVD} Diamond Detectors: {Position} and Temporal Response Measurements using a Synchrotron Microbeam Probe}, 
  volume={1039}, 
  DOI={10.1557/PROC-1039-P06-02}, 
  journal={MRS OPL}, 
  publisher={Cambridge University Press}, 
  author={Morse, J. and Salom\'e, M. and Berdermann, E. and Pomorski, M. and Grant, J. and O'Shea, V. and Ilinski, P.}, 
  year={2007}, 
  pages={1039–P06-02}}

@article{keister2018,
author = {Keister, J. W. and Cibik, L. and Schreiber, S. and Krumrey, M.},
title = "{Characterization of a quadrant diamond transmission X-ray detector including a precise determination of the mean electron{--}hole pair creation energy}",
journal = "J. Synchrotron Radiat.",
year = "2018",
volume = "25",
number = "2",
pages = "407--412",
doi = {10.1107/S1600577517017659}
}

@article{kobayashi1972,
    author = {Kobayashi, T.},
    title = {Average energy to form electron‐hole pairs in {GaP} diodes with alpha particles},
    journal = {Appl. Phys. Lett.},
    volume = {21},
    number = {4},
    pages = {150-152},
    year = {1972},
    month = {08},
    doi = {10.1063/1.1654321}
    }

@article{bertuccio2002,
    author = {Bertuccio, G. and Maiocchi, D.},
    title = {Electron-hole pair generation energy in gallium arsenide by $x$ and $\gamma$ photons},
    journal = {J. Appl. Phys.},
    volume = {92},
    number = {3},
    pages = {1248-1255},
    year = {2002},
    month = {08},
    doi = {10.1063/1.1490158}
    }

@article{lowe2007,
title = {A measurement of the electron–hole pair creation energy and the {Fano} factor in silicon for 5.9 {keV X-rays} and their temperature dependence in the range 80-270 K},
journal = {Nucl. Instrum. Methods Phys. Res. A},
volume = {576},
number = {2},
pages = {367-370},
year = {2007},
month = {06},
doi = {https://doi.org/10.1016/j.nima.2007.03.020},
author = {B.G. Lowe and R.A. Sareen}
}

@article{scholze2000,
title = {Determination of the electron–hole pair creation energy for semiconductors from the spectral responsivity of photodiodes},
journal = {Nucl. Instrum. Methods Phys. Res. A},
volume = {439},
number = {2},
pages = {208-215},
year = {2000},
month = {01},
doi = {https://doi.org/10.1016/S0168-9002(99)00937-7},
author = {F. Scholze and H. Henneken and P. Kuschnerus and H. Rabus and M. Richter and G. Ulm},
}

@article{sangalli2016,
  title = {Nonequilibrium optical properties in semiconductors from first principles: {A} combined theoretical and experimental study of bulk silicon},
  author = {Sangalli, D. and Dal Conte, S. and Manzoni, C. and Cerullo, G. and Marini, A.},
  journal = {Phys. Rev. B},
  volume = {93},
  issue = {19},
  pages = {195205},
  numpages = {13},
  year = {2016},
  month = {05},
  doi = {10.1103/PhysRevB.93.195205}
}

@article{shishkin2007,
  title = {Self-consistent {GW} calculations for semiconductors and insulators},
  author = {Shishkin, M. and Kresse, G.},
  journal = {Phys. Rev. B},
  volume = {75},
  issue = {23},
  pages = {235102},
  year = {2007},
  month = {06},
  doi = {https://doi.org/10.1103/PhysRevB.75.235102}
}

@article{chen2015,
  title = {Accurate band gaps of extended systems via efficient vertex corrections in $GW$},
  author = {Chen, W. and Pasquarello, A.},
  journal = {Phys. Rev. B},
  volume = {92},
  issue = {4},
  pages = {041115(R)},
  numpages = {5},
  year = {2015},
  month = {07},
  doi = {10.1103/PhysRevB.92.041115},
}

@article{bauer1994,
  title = {Direct Determination of Impact Ionization Quantum Yield in {Si} by Ballistic-Electron-Emission Microscopy},
  author = {Bauer, A. and Ledeke, R.},
  journal = {Phys. Rev. Lett.},
  volume = {72},
  issue = {6},
  pages = {928--931},
  year = {1994},
  month = {02},
  doi = {https://doi.org/10.1103/PhysRevLett.72.928},
}

@article{fischetti2019,
    author = {Fischetti, M. V. and Yoder, P. D. and Khatami, M. M. and Gaddemane, G. and Van de Put, M. L.},
    title = {{``Hot electrons in Si lose energy mostly to optical phonons'': Truth or myth?}},
    journal = {Appl. Phys. Lett.},
    volume = {114},
    number = {22},
    pages = {222104},
    year = {2019},
    month = {06},
    doi = {10.1063/1.5099914},
}

\end{document}